\documentstyle[eqsecnum,aps,prd,epsf]{revtex}

\def\msol{\ifmmode M_\odot\else$M_\odot$\fi}
\newcommand{\deq}{{\stackrel{\cdot}{=}}}

\begin{document}

% UTB Logo
%\vspace*{-1.5cm}
%\epsfxsize=1.1cm
%\leftline{\epsfbox{UTB_x.eps}}
%\vspace{-1.5cm}
%{
%\baselineskip20pt
%\hspace{1cm} \hbox{\large
%THE {\Large U}NIVERSITY OF {\Large T}EXAS AT {\Large B}ROWNSVILLE} \par
%\hspace{1cm} \hbox{\large
%{\Large T}EXAS {\Large S}OUTHMOST {\Large C}OLLEGE} \par
%}
%\vspace{2mm}
% UTB Logo end

% CGWA Logo
%\epsfxsize=1.5cm
%\leftline{\epsfbox{CGWA_x.eps}}
%\vspace{-0.5cm}
%\hspace{1.1cm}{\Large\sl The Center for Gravitational Wave Astronomy}
%\vspace{3mm}
% CGWA Logo end

% CalTech Logo
%\hspace{-10mm}
%\leftline{\epsfbox{caltech_logo.eps}}
%\vspace{-10.0mm} % for revtex
%\thispagestyle{empty}
%{\baselineskip-4pt
%\font\yitp=cmmib10 scaled\magstep2
%\font\elevenmib=cmmib10 scaled\magstep1  \skewchar\elevenmib='177
%\leftline{\baselineskip20pt
%\hspace{25mm} % for revtex
%\hspace{-100mm} % for article
%\vbox to0pt
%   { {\yitp\hbox{California \hspace{1.5mm} Institute \hspace{1.5mm}
%of \hspace{1.5mm} Technology} }
%     {\large\sl\hbox{{TAPIR}} }\vss}}
% CalTech Logo end

%\begin{figure}
%\leftline{\includegraphics{caltech_logo}}
%\nonumber
%\end{figure}
%\font\yitp=cmmib10 scaled\magstep2
%\font\elevenmib=cmmib10 scaled\magstep1  \skewchar\elevenmib='177
%\leftline{\baselineskip20pt
%\hspace{3.5cm}
%\vbox to0pt
%   { {\yitp\hbox{California \hspace{1.5mm} Institute \hspace{1.5mm}
%of \hspace{1.5mm} Technology} }
%     {\large\sl\hbox{{TAPIR}} }\vss}}
%\vspace{0.5cm}

%preprint number
%\begin{flushright}
%
%\end{flushright}
%preprint number end

\bigskip
\bigskip
\bigskip

\begin{center}

{\Large \bf
From the self-force problem to the Radiation reaction formula
}

\bigskip

{\large Yasushi Mino
\footnote{Electronic address: mino@tapir.caltech.edu}}\\

\medskip

{\em mail code 130-33 Caltech Pasadena CA 91125 USA}\\

\medskip

\today

\end{center}

\bigskip

{\bf Abstract} \\

We review a recent theoretical progress
in the so-called self-force problem 
of a general relativistic two-body system.
Although a two-body system in Newtonian gravity is a very simple problem,
some fundamental issues are involved in relativistic gravity.
Besides, because of recent projects for gravitational wave detection,
it comes to be possible to see those phenomena
directly via gravitational waves,
and the self-force problem becomes
one of urgent and highly-motivated problems in general relativity.
Roughly speaking, there are two approaches to investigate this problem;
the so-called post-Newtonian approximation, 
and a black hole perturbation.

In this paper, we review a theoretical progress
in the self-force problem using a black hole perturbation.
Although the self-force problem seems 
to be just a problem to calculate a self-force, 
we discuss that the real problem is 
to define a gauge invariant concept of a motion 
in a gauge dependent metric perturbation. 

%%%%%%%%%%%%%%%%%%%%%%%%%%%%%%%%%%%%%%%%%%%%%%%%%%%%%%%%%%%
%%%%%%%%%%%%%%%%%%%%%%%%%%%%%%%%%%%%%%%%%%%%%%%%%%%%%%%%%%%
\section{Introduction} \label{sec:intro}
%%%%%%%%%%%%%%%%%%%%%%%%%%%%%%%%%%%%%%%%%%%%%%%%%%%%%%%%%%%

For a single isolated body, the self-induced gravitational field 
does not accelerate the body
because of the momentum conservation.
However, in a multi-body system in general relativity,
there could be the case that
the self-induced gravitational field could accelerate the body.
Although one may naively say we have the self-force acting on the body, 
the concept of the body's motion is itself ambiguous in general relativity
because one can freely choose a reference coordinate system
to describe it.
Thus, even if we could derive an explicit form of the self-force,
there still remains a fundamental question;
what we learn from the the self-force about the body's motion.
Here we focus on a binary system
as a simple example to consider this fundamental question.

The dynamics of a binary system is also motivated 
by rapidly-progressing projects of gravitational wave detection.
The so-called post-Newtonian approximation has a fairly long history
in calculating an evolution of a equal-mass binary system.
Although we now have a huge amount of interesting results,
the post-Newtonian approximation was usually done
only with a few choices of the coordinate system.

Because the choice of coordinate systems is
a key issue in understanding this fundamental question, 
it would be more interesting to consider another approach 
in which we may have a more degree of freedom to choose coordinate systems. 
Here we consider the self-force problem 
by a black hole perturbation. 
We consider the relativistic two body problem 
as a background black hole and a point particle 
orbiting around the black hole. 
We assume that the background black hole 
is described by a Kerr geometry, 
and consider a metric perturbation 
induced by a particle moving along a general bound orbit. 
The reference coordinate system for the particle's motion 
is dominantly given 
by the coordinate system of the background Kerr geometry, 
but, since the background geometry is perturbed 
by the self-gravity of the particle, 
we have a small freedom to choose the coordinate system. 
This is what we know 
as a gauge freedom of the metric perturbation. 

This two-body system is considered to be a good approximation 
of a gravitational-wave source we may see by the LISA project. 
Because of the long base line, 
LISA will see low frequency gravitational waves 
(around $10$mHz to $0.1$Hz).
Gravitational waves in this frequency range 
may be obtained from an orbit around a supermassive black hole 
(around $10^6$ to $10^8$ \msol), 
which we may astrophysically expect 
to exist at the center of a galaxy. 
Our recent understanding of a dynamical structure of a galactic core 
suggests that supermassive black holes in the observable range of LISA 
would capture stellar mass compact objects 
in their gravitational potentials. 
Those gravitationally captured objects are 
inspiralling around the black holes, 
emitting gravitational waves detectable by LISA.

%%%%%%%%%%%%%%%%%%%%%%%%%%%%%%%%%%%%%%%%%%%%%%%%%%%%%%%%%%%
%%%%%%%%%%%%%%%%%%%%%%%%%%%%%%%%%%%%%%%%%%%%%%%%%%%%%%%%%%%
\section{Overview} \label{sec:ov}
%%%%%%%%%%%%%%%%%%%%%%%%%%%%%%%%%%%%%%%%%%%%%%%%%%%%%%%%%%%

The study of an orbital motion of a particle in a black hole geometry 
has started from the discovery 
of the so-called MiSaTaQuWa self-force\cite{sf}, 
which we shall review in Sec.\ref{sec:sf}. 
At this moment, the question was how we deal with a point particle 
and we obtained a general covariant expression of the self-force 
in a given vacuum background. 
Since the MiSaTaQuWa self-force is just a formal result,
our next question was to evaluate the self-force 
explicitly in a Kerr geometry, 
using a known technique to calculate a metric perturbation. 
We review some basic ideas of 
some promising calculation methods in Sec.\ref{sec:reg}. 

There are two approaches. 
One is to calculate the self-force directly 
by explicitly implementing the point splitting regularization
of the MiSaTaQuWa self-force, 
and we call it the self-force calculation. 
The other is to calculate 
the radiation reaction to the constants of motion 
based on our understanding of the self-force, 
and we call it the radiation reaction formula. 
By the self-force calculation, 
one may have the exact self-force 
in the framework of a linear perturbation, 
however, we still have a technical problem 
when we consider the self-force calculation in a Kerr black hole. 
Besides, as we shall discuss in Sec.\ref{sec:sf?}, 
there is a theoretical issue 
whether the self-force derived by this method 
correctly describes the radiation reaction effect. 
On the other hand, the radiation reaction formula 
considers radiation reaction to the orbit. 
Different from the self-force calculation, 
we only calculate the radiative part of the self-force 
in a class of gauge conditions where we are guaranteed to include 
the radiation reaction effect to the orbit. 
It has a technical advantage that 
the calculation method of this formula is already established. 
Furthermore, our recent result shows that 
the radiation reaction formula may have 
the most optimal prediction of the orbit. 

It is widely believed that 
the self-force describes the orbital evolution 
with the effect of gravitational radiation reaction. 
Sec.\ref{sec:sf?} discusses that this is not true in general. 
We point out that a usual scheme of a metric perturbation 
describes a binary evolution only for a finite time interval. 
We show that the self-force could vanish by a special gauge transformation 
for this whole time interval, irrespective to gravitational radiation 
we see from the asymptotic gravitational wave flux. 
Thus, strictly speaking, 
the self-force has nothing to do with gravitational radiation reaction 
in the usual metric perturbation 
without some restriction of gauge conditions. 
We consider this problem happens 
not because of our definition of the self-force 
but because the usual scheme of a metric perturbation 
cannot describe the binary evolution long enough. 
In order to solve this, 
we propose an adiabatic extension of a metric perturbation, 
which we review in Sec.\ref{sec:ad}. 
The adiabatic extension is possible 
under a class of gauge conditions, 
where the self-force is guaranteed to include 
the radiation reaction effect. 
We also discuss the validity of 
the adiabatic approximation of the linear metric perturbation 
and the self-force defined by this. 
We find that the self-force in this class of gauge conditions 
might not predict the binary evolution sufficiently long 
for the LISA project in general. 
In order to extend the validity, 
we introduce a special gauge condition 
where the radiation reaction formula gives an exact self-force. 
In this gauge condition, we find that 
the self-force may predict the binary evolution 
long enough for the LISA project.

%%%%%%%%%%%%%%%%%%%%%%%%%%%%%%%%%%%%%%%%%%%%%%%%%%%%%%%%%%%
%%%%%%%%%%%%%%%%%%%%%%%%%%%%%%%%%%%%%%%%%%%%%%%%%%%%%%%%%%%
\section{self-force} \label{sec:sf}
%%%%%%%%%%%%%%%%%%%%%%%%%%%%%%%%%%%%%%%%%%%%%%%%%%%%%%%%%%%

In this section, we review 
a derivation of the MiSaTaQuWa self-force. 
In this review, we focus on the physical motivation 
why we take our calculation strategy 
and we omit technical details of the calculation. 
We suggest the readers interested in those details 
to refer our original paper, or, 
the review papers by us or Poisson\cite{sf}. 

%%%%%%%%%%%%%%%%%%%%%%%%%%%%%%%%%%%%%%%%%%%%%%%%%%%%%%%%%%%
\subsection{A historical motivation} 
%%%%%%%%%%%%%%%%%%%%%%%%%%%%%%%%%%%%%%%%%%%%%%%%%%%%%%%%%%%

In a black hole perturbation, 
a point particle is not well-defined 
because the self-gravity of the point particle diverges 
and the metric perturbation becomes invalid around the particle. 
Nevertheless, we had been using the point particle 
without any rigorous supporting argument 
because of the technical simplicity 
and because the point particle might be 
an only possible description of a star 
whose internal structure is supposed to be unimportant, 
and, a metric perturbation induced by a point particle 
has been investigated for a long time\cite{pp} 
before the MiSaTaQuWa self-force was derived\cite{sf}. 

At this time, we used the so-called Press-Teukolsky formalism\cite{PT} 
which gives us a finite energy and angular momentum flux 
carried away by gravitational waves. 
Even if the point particle induces 
the divergent gravitational perturbation, 
one can define a finite energy and angular momentum 
on a $t$-constant hypersurface in the background Kerr geometry
and we consider that the loss of the energy and angular momentum 
by the gravitational wave emission 
describes radiation reaction 
to the particle's motion. 
This method is now refered by the balance formula. 

The balance formula was considered to be successful 
to describe radiation reaction to a particle 
moving along a quasi-circular orbit or an equatorial orbit. 
In this case, we consider an adiabatic approximation of the orbit 
where the orbit is approximated by a geodesic at each instant. 
Then, instead of the evolution of the orbital coordinates, 
we consider the evolution of "$3$ constants of motion" 
(the orbital energy $E$, $z$ component of the angular momentum $L$ 
and Carter constant $C$) of geodesics around a Kerr geometry. 
The balance formula tells us radiation reaction 
to only two of these constants, $E$ and $L$, 
and the problem was to derive radiation reaction to the rest, $C$. 
When the orbit is either quasi-circular or equatorial, 
one can prove that it will remain quasi-circular or equatorial 
under radiation reaction\cite{rrb}, 
as a result, we have an algebraic relation with $E$, $L$ and $C$ 
to derive radiation reaction to $C$ by the balance formula. 

The difficulty to compute radiation reaction to $C$ 
was that Carter constant is not associated with the Killing vector. 
The Press-Teukolsky formalism was derived 
because $E$ and $L$ are the conserved currents 
defined on a $t$-constant hypersurfaces
in association with the symmetry of the background spacetime. 
On the other hand, Carter constant is defined only along a geodesic 
and has nothing to do with the symmetry. 
Besides, it came to be aware that, 
in order to integrate the geodesic equation, 
one needs three phase constants 
in addition to the constants of motion, 
and we did not know how those phase constants evolve 
by gravitational radiation reaction. 

At this moment, a simplest solution of this problem 
seemed to be an explicit calculation of the self-force. 
We considered that the self-force must have contained 
all the information of the orbit with the effect of radiation reaction, 
and that, after computing the orbit using the self-force, 
one can calculate a gravitational waveform 
by a linear metric perturbation formalism. 
As we shall argue in Sec.\ref{sec:sf?}, 
this statement is proven to be incorrect 
because a usual perturbation formalism cannot describe 
radiation reaction to the orbit, 
nevertheless, this idea had been spread widely 
in the gravitational wave community. 

%%%%%%%%%%%%%%%%%%%%%%%%%%%%%%%%%%%%%%%%%%%%%%%%%%%%%%%%%%%
\subsection{Problem of Divergence} 
%%%%%%%%%%%%%%%%%%%%%%%%%%%%%%%%%%%%%%%%%%%%%%%%%%%%%%%%%%%

A point particle is defined by the Dirac's delta function. 
It has no volume, but, a finite mass. 
This makes a divergent metric perturbation 
at the location of the particle. 
Because of the divergence, we have two problems; \\ 
{\bf 1)} 
{\it Can we use the point particle to induce a linear metric perturbation, 
which invalidates the linear perturbation scheme?} \\
{\bf 2)} 
{\it Can we extract a finite physical information of a self-force 
from the divergent field?} \\ 

A common idea to understand a divergent field 
in Quantum Field Theory is renormalization. 
In electromagnetism, a point charge induces 
a divergent vector potential at the location of the particle. 
When the charge moves in a curved spacetime, 
this self-induced vector potential will act on the charge's motion. 
Because the quantum electrodynamics is renormalizable, 
a formal derivation of this electromagnetic self-force in a curved spacetime 
was successfully made by a mass renormalization\cite{DB}. 
Here one evaluates a surface integration of the stress-energy tensor 
of the charge and the electromagnetic field 
over a world tube surrounding the orbit. 
The sum of the surface integral must vanish 
because of the gauss's law,
and we obtain the self-force equation 
including the effect of electromagnetic radiation reaction. 

We considered an analogous calculation for a gravitational self-force 
by the mass renormalization\cite{sf}. 
We do not have the stress-energy tensor 
of the particle and the full gravitational field, 
however, we could find a conserved current 
of the particle and the metric perturbation 
in the background geometry. 
We consider to expand the Einstein tensor 
by the metric perturbation $h_{\mu\nu}$ in the background $g_{\mu\nu}$ as 
\begin{eqnarray}
G_{\mu\nu}[g+h] &=& 
G_{\mu\nu}[g]+G^{(1)}_{\mu\nu}[h]+G^{(2+)}_{\mu\nu}[h] \,, 
\end{eqnarray}
where $G^{(1)}_{\mu\nu}$ and $G^{(2+)}_{\mu\nu}$ represent 
the linear terms and all the non-linear terms of the expansion 
with respect to $h_{\mu\nu}$. 
Because we assume the background is a vacuum solution, 
we have $G_{\mu\nu}[g]=0$. 
It is well-known that the divergence of $G^{(1)}_{\mu\nu}[h]$ 
with respect the background geometry vanishes algebraically. 
Thus, we have the conserved current in the background geometry as 
\begin{eqnarray}
{\cal T}_{\mu\nu} &=& G^{(2+)}_{\mu\nu}[h] +T_{\mu\nu} \,,
\end{eqnarray}
where $T_{\mu\nu}$ is the stress-energy tensor of the particle. 
Because $G^{(2+)}_{\mu\nu}[h]$ is quadratic in $h_{\mu\nu}$ 
to the leading order of the metric perturbation, 
we may consider this as an "effective stress-energy tensor" 
of the particle and the metric perturbation 
defined in the background geometry. 
By doing a surface integration over a world tube, 
we obtained a gravitational self-force equation 
by the mass renormalization. 

Although we could have derived the MiSaTaQuWa self-force 
by the mass renormalization, 
the derivation did not answer the above fundamental questions. 
A crucial issue in this derivation is that 
the perturbative quantum gravity is known nonrenormalizable, 
thus, there is no theoretical support to use 
the mass renormalization in deriving the gravitational self-force. 
Furthermore, a physical interpretation of the motion 
is not clear yet. 
If we suppose that 
the "point particle" represents a black hole, for example, 
the center of the particle is inside the event horizon, 
and one may ask 
what it means to consider a motion hidden inside the horizon. 

%%%%%%%%%%%%%%%%%%%%%%%%%%%%%%%%%%%%%%%%%%%%%%%%%%%%%%%%%%%
\subsection{Matched Asymptotic Expansion} \label{ssec:mae}
%%%%%%%%%%%%%%%%%%%%%%%%%%%%%%%%%%%%%%%%%%%%%%%%%%%%%%%%%%%

In order to answer these fundamental questions 
on the use of the "point particle", 
we consider another approach. 
Since the primary problem here is 
a theoretical justification to use the "point particle" 
rather than just a calculation of the self-force, 
we consider to construct a metric itself 
by using a matched asymptotic expansion technique. 
By the matched asymptotic expansion, 
we consider to locate a small black hole of a length scale $m$ 
in a background spacetime of a curvature scale $L$. 
Assuming that $m<<L$, one could construct the metric 
perturbatively by a small parameter $m/L<<1$. 
This approach was originally taken 
by the post-Newtonian calculation\cite{match}, 
where we have an explicit expression of a metric easy to control. 
In this problem, we consider to locate the small black hole 
in a general vacuum background spacetime 
and, in order to have an explicit expression of the metric, 
we use the so-called bi-tensor formalism 
introduced in Ref.\cite{DB}.

With a radial distance $r(>0)$ from the center of the small black hole, 
the gravitational field is determined dominantly 
by that of the small black hole at $r<L$, 
and by the background spacetime at $r>m$. 
Hence, the metric at $r<L$ is constructed 
by the small black hole with a black hole perturbation 
(we call the near-zone expansion) as 
\begin{eqnarray}
g^{near}_{\mu\nu} &=& g^{BH}_{\mu\nu}+L^{-1}g^{(1/)}_{\mu\nu}
+L^{-2}g^{(2/)}_{\mu\nu}+\cdots \,, \label{eq:nz}
\end{eqnarray}
where $g^{BH}_{\mu\nu}$ is the metric of the small black hole. 
The black hole perturbation $g^{(n/)}_{\mu\nu}$ is defined 
with an appropriate boundary condition 
on the black hole horizon. 
As we cannot specify a boundary condition 
for $g^{(n/)}_{\mu\nu}$ at $r \geq L$, 
the metric perturbation cannot be determined uniquely 
until we complete the matching. 
The metric at $r>m$ is constructed 
by the background spacetime with its perturbation 
(we call the far-zone expansion) as 
\begin{eqnarray}
g^{far}_{\mu\nu} &=& g^{bg}_{\mu\nu}+m g^{[/1]}_{\mu\nu}
+m^2 g^{[/2]}_{\mu\nu}+\cdots \,, \label{eq:fz} 
\end{eqnarray}
where $g^{bg}_{\mu\nu}$ is the metric of the background spacetime. 
The perturbation of the background $g^{[/n]}_{\mu\nu}$ is 
derived with an appropriate asymptotic boundary condition 
and is not determined uniquely 
because of the lack of a boundary condition at $r \leq m$. 
Instead of the boundary condition at $r \leq m$, 
one can put an auxiliary matter at $r \leq m$ 
to induce a vacuum perturbation at $r>m$, 
and we use a point source with an arbitrary internal structure
derived in Ref.\cite{Di}. 

The matched asymptotic expansion can be done 
at the radius where both near-zone and far-zone expansions are valid, 
which we call the overlapping zone. 
Suppose we take the radius of the overlapping zone 
as $r\sim\sqrt{m L}$, 
and one can re-expand (\ref{eq:nz}) and (\ref{eq:fz}) as 
\begin{eqnarray}
g^{near}_{\mu\nu} &=& 
(g^{(0/0)}_{\mu\nu}+m g^{(0/1)}_{\mu\nu}+\cdots)
+L^{-1}(g^{(1/0)}_{\mu\nu}+m g^{(1/1)}_{\mu\nu}+\cdots)
+\cdots \,, \label{eq:nfz} \\ 
g^{far}_{\mu\nu} &=& 
(g^{[0/0]}_{\mu\nu}+L^{-1}g^{[1/0]}_{\mu\nu}+\cdots)
+m(g^{[0/1]}_{\mu\nu}+L^{-1}g^{[1/1]}_{\mu\nu}+\cdots)
+\cdots \,. \label{eq:fnz} \\ 
\end{eqnarray}
From a dimensional analysis, we may see 
$g^{(n/m)}_{\mu\nu}\sim g^{[n/m]}_{\mu\nu} \sim r^{n-m}$, 
thus, at the overlapping zone, one can estimate the expansions as 
\begin{eqnarray}
L^{-n}m^m g^{(n/m)}\sim L^{-n}m^m g^{[n/m]}\sim 
(m/L)^{(n+m)/2} \,. 
\end{eqnarray}
Under the matched asymptotic expansion, 
we assume $m/L<<1$ and we equate 
$g^{(n/m)}_{\mu\nu}=g^{[n/m]}_{\mu\nu}$ one by one 
from small to large $n+m$ 
which will perturbatively determine 
the boundary condition of (\ref{eq:nz}) at $r \geq L$ 
and the internal structure of the auxiliary matter 
to induce the perturbation of (\ref{eq:fz}). 

Here, it is important to note coordinate conditions. 
Since the derivation of (\ref{eq:nz}) and (\ref{eq:fz}) 
are totally independent, 
they may use different coordinate systems, 
and we have to introduce a coordinate transformation 
so that we can correctly evaluate 
$g^{(n/m)}_{\mu\nu}=g^{[n/m]}_{\mu\nu}$. 
It is also important to fix a gauge condition 
for the metric of the small black hole 
in order to fix the location of the particle 
in the near-zone expansion. 

We suppose 
that the background metric and its coordinate system 
are given by $(g^{bg}_{\alpha\beta},\{x^\alpha\})$ 
and that the orbit in the background metric is 
described by $z^\alpha(\tau)$ with a parameter $\tau$. 
We also suppose 
that the black hole metric and its coordinate system 
are given by $(g^{BH}_{ij},\{X^i,i=0,1,2,3\})$ 
and we take a Schwarzschild black hole metric 
in the harmonic coordinates 
where $X^0$ is the temporal coordinate 
and $X^1,X^2,X^3$ are spatial. 
In order to fix the location of the small black hole 
in the near-zone expansion, 
we take the gauge condition such that 
the $l=1$ modes of the perturbation $L^{-n} g^{(n/)}_{ij}$ 
in the tensor-spherical decomposition vanish. 
Then, we may say the particle is located at $X^1=X^2=X^3=0$ 
in the near-zone expansion. 

For the matched asymptotic expansion of the metrices 
in the far-zone expansion and the near-zone expansion, 
we define the coordinate transformation 
by using a tetrad defined along the orbit 
and derivatives of the bi-scalar of 
the half-the-squared geodesic distance. 
The coordinate transformation is 
determined perturbatively 
by the matched asymptotic expansion of the metrices. 

%%%%%%%%%%%%%%%%%%%%%%%%%%%%%%%%%%%%%%%%%%%%%%%%%%%%%%%%%%%
\subsection{MiSaTaQuWa Self-force} 
%%%%%%%%%%%%%%%%%%%%%%%%%%%%%%%%%%%%%%%%%%%%%%%%%%%%%%%%%%%

We find that the coordinate transformation gives us 
the information of the particle's orbit because it relates 
the location of the small black hole in the near-zone expansion 
(i.e. $X^1=X^2=X^3=0$) 
to the coordinates in the far-zone expansion. 
By the matching $g^{(0/0)}_{\mu\nu}=g^{[0/0]}_{\mu\nu}$ 
and $g^{(1/0)}_{\mu\nu}=g^{[1/0]}_{\mu\nu}$, 
we find that the particle moves along a geodesic. 
By the matching 
$g^{(2/0)}_{\mu\nu}=g^{[2/0]}_{\mu\nu}$, 
$g^{(0/1)}_{\mu\nu}=g^{[0/1]}_{\mu\nu}$, 
$g^{(1/1)}_{\mu\nu}=g^{[1/1]}_{\mu\nu}$ 
and $g^{(2/1)}_{\mu\nu}=g^{[2/1]}_{\mu\nu}$, 
we derive the self-force. 

The point particle is shown to be consistent 
for the derivation of the metric in the far-zone expansion 
as a result of the matched asymptotic expansion of the metric. 
Even though the point particle makes a divergent metric perturbation, 
it is just an auxiliary source 
to induce a finite metric perturbation at $r>m$ 
in the far-zone expansion. 
This justifies the use of a point particle 
in calculating an energy and angular momentum flux 
via an asymptotic metric perturbation 
since it uses the metric perturbation only in the far-zone expansion. 

We also consider that the derivation of the self-force 
by the matched asymptotic expansion 
suggests a physical interpretation of the particle's motion. 
In the end, the motion of the particle is 
considered by the coordinate system in the far-zone expansion 
which is the background metric plus a small perturbation 
induced by the small black hole. 
For an observer far from the particle, 
the background metric dominantly determines a reference frame 
to measure the motion of the particle with, 
whatever non-perturbative things happen 
around the small black hole, 
thus, the self-force derived by the matched asymptotic expansion 
is expected to be physically meaningful for such an observer.
Because the small black hole is small 
compared to the curvature length of the background metric, 
one can assign the location of the particle 
in the coordinate system of the regular background metric, 
and consider the motion of the particle without divergence. 
Here we have a small ambiguity in this assignment of the particle's location 
because of a finite size of the small black hole, 
which can be interpreted as a gauge ambiguity of the self-force. 
The location of the black hole in the near-zone metric 
is determined by the gauge condition of the metric perturbation 
$L^{-n} g^{(n/)}_{\alpha\beta}$. 
In the present derivation of the self-force, 
the gauge conditions of $L^{-1} g^{(1/1)}_{}$ 
and $L^{-2} g^{(2/1)}_{ij}$ are important. 
By the matched asymptotic expansion, 
they fix the gauge condition of the linear metric perturbation 
$m g^{[/1]}_{\alpha\beta}$ of the far-zone expansion.

%%%%%%%%%%%%%%%%%%%%%%%%%%%%%%%%%%%%%%%%%%%%%%%%%%%%%%%%%%%
%%%%%%%%%%%%%%%%%%%%%%%%%%%%%%%%%%%%%%%%%%%%%%%%%%%%%%%%%%%
\section{Regularization Calculation} \label{sec:reg}
%%%%%%%%%%%%%%%%%%%%%%%%%%%%%%%%%%%%%%%%%%%%%%%%%%%%%%%%%%%

Since the MiSaTaQuWa self-force is just a formal expression 
of a self-force in an arbitrary background, 
the next problem was to explicitly calculate a self-force 
in a specific background, especially, a Kerr black hole 
for an application to LISA project. 
The formal result of the self-force\cite{sf} can be written as 
\begin{eqnarray}
F^\alpha(\tau) &=& \lim_{x\to z(\tau)}F^\alpha[h^{reg}(x)] 
\label{eq:sf} \,, \\ 
F^\alpha[k(x)] &=& 
\delta\Gamma^\alpha_{\beta\gamma}[k(x)]v^\beta v^\gamma
+v^\alpha \delta\Gamma_{\beta\gamma\delta}[k(x)]
v^\beta v^\gamma v^\delta \,, \\ 
h^{reg}_{\alpha\beta}(x) &=& 
h_{\alpha\beta}(x)-h^{sing}_{\alpha\beta}(x) \,, \quad 
(x^\alpha \not = z^\alpha(\tau)) \,, 
\end{eqnarray}
where $x^\alpha$ is a field point, $z^\alpha(\tau)$ is an orbit, 
and $v^\alpha$ is the $4$-velocity. 
$\delta\Gamma^\alpha_{\beta\gamma}$ is 
a linear tensor differential operator acting on a metric perturbation 
to calculate a perturbed Christoffel symbol. 
$h_{\alpha\beta}(x)$ is a linear metric perturbation 
induced by a point particle, 
and $h^{sing}_{\alpha\beta}(x)$ is a singular part 
derived by the Hadamard expansion. 
$h^{reg}_{\alpha\beta}(x)$ is called a regularized metric perturbation. 
Both $h_{\alpha\beta}(x)$ and $h^{sing}_{\alpha\beta}(x)$
are divergent along the orbit, 
but, $h^{reg}_{\alpha\beta}(x)$ is regular along the orbit 
since the divergence 
of $h_{\alpha\beta}(x)$ and $h^{sing}_{\alpha\beta}(x)$ 
cancel each others. 
Thus, the calculation of the self-force can be done 
by first evaluating it at a field point $x^\alpha$, 
then taking the limit $x^\alpha \to z^\alpha(\tau)$. 
By doing this point splitting regularization, we have a finite result. 

There are two promising approaches. 
One is to calculate the self-force directly 
by doing a Harmonic mode decomposition, 
and we call it the self-force calculation. 
The other is to calculate 
radiation reaction to the constants of motion 
based on our theoretical understanding of the self-force, 
and we call it the radiation reaction formula. 

%%%%%%%%%%%%%%%%%%%%%%%%%%%%%%%%%%%%%%%%%%%%%%%%%%%%%%%%%%%
\subsection{Self-Force Calculation} \label{ssec:sfc}
%%%%%%%%%%%%%%%%%%%%%%%%%%%%%%%%%%%%%%%%%%%%%%%%%%%%%%%%%%%

Here we briefly review a method to compute a self-force directly. 
The basic idea here is that, 
since it is difficult to compute 
$F^\alpha[h^{reg}(x)]$ of (\ref{eq:sf}) directly, 
we consider to calculate $F^\alpha[h(x)]$ and $F^\alpha[h^{sing}(x)]$, 
then, the self-force can be derived by 
\begin{eqnarray}
F^\alpha(\tau) &=& \lim_{x\to z(\tau)}
(F^\alpha[h(x)]-F^\alpha[h^{sing}(x)]) \,. \label{eq:sff}
\end{eqnarray}
We call $F^\alpha[h(x)]$ and $F^\alpha[h^{sing}(x)]$ 
by a bare self-force field and a singular self-force field, 
which is defined at a field point $x^\alpha$ 
not on the orbit $z^\alpha(\tau)$. 
It is important to note that 
the subtraction of (\ref{eq:sff}) must be done 
before taking the coincidence limit $x^\alpha\to z^\alpha(\tau)$ 
since $F^\alpha[h(x)]$ and $F^\alpha[h^{sing}(x)]$ 
are divergent along the orbit. 

This type of calculation is called regularization. 
Various techniques for regularization calculations 
are well-known in quantum field theory, 
but we cannot use those techniques because of two reasons; \\ 
{\bf 1)}
{\it In a usual regularization of quantum field theory, 
we consider a theory of a global Lorentz invariance 
and we can expand the divergent field by the Fourier mode, 
respecting the Lorentz symmetry. 
However, a black hole spacetime does not have such a symmetry 
and we may only decompose $h_{\alpha\beta}(x)$ 
into Fourier-Harmonic series.}\\ 
{\bf 2)}
{\it $h^{sing}_{\alpha\beta}(x)$ is derived by the Hadamard expansion, 
thus, we have its local divergent expression only around the orbit.} \\
Besides, in the case of a Kerr black hole as a background, 
a method to calculate a linear metric perturbation was proposed 
just recently in Ref.\cite{amos}, 
and it has not been coded yet. 
Another simple technique to calculate a metric perturbation 
was proposed in Ref.\cite{chrz}, 
however, it is known to be valid 
only for a homogeneous metric perturbation 
and it cannot describe a metric perturbation around the orbit. 
It is also important to note that 
the original self-force was derived in the harmonic gauge condition, 
thus, we may have a residual divergence by a difference of gauge conditions 
unless we calculate $h_{\alpha\beta}(x)$ 
in the harmonic gauge condition. 

The most successful method at present is 
the so-called mode decomposition regularization\cite{self}, 
which is actually based on two ideas. 
One is to use a local coordinate expansion 
originally proposed by us\cite{pow}, 
and the other is to use the so-called regularization parameters 
originally proposed by Barack, Burko and Ori\cite{par}. 
Here we briefly review 
a basic idea of this regularization calculation 
in the case that the background is a Schwarzschild black hole. 
There is also some progress on an extension to a Kerr black hole 
and a method to deal 
with the residual divergence by a gauge condition, 
but, we do not discuss it here because of its technicality. 

We use the Schwarzschild coordinates 
and denote a field point and an orbital point 
by $x^\alpha=\{t,r,\theta,\phi\}$ 
and $z^\alpha=\{t_0,r_0,\theta_0,\phi_0\}$. 
We suppose that a full metric perturbation is derived
by a harmonic decomposition 
and we calculate a harmonic decomposition of 
the bare self-force field.\footnote{
Since we are interested in calculating a self-force vector, 
there are some options in the harmonic decomposition. 
In order to avoid a complication of multiple components, 
we consider it easier to decompose 
each component of the self-force vector by the scalar harmonics.
For this technical detail, see Ref.\cite{self}.} 

After the harmonic decomposition, 
the expansion coefficients become finite functions of $(t,r)$, 
and the divergence of the bare self-force field appears 
only by taking the infinite sum over harmonics 
at the coincidence limit $x^\alpha \to z^\alpha$. 
We take the angular coincidence limit $(\theta,\phi)\to(\theta_0,\phi_0)$, 
and take the sum over the azimuthal modes. 
Then the mode decomposition of the bare self-force field 
can schematically be written as 
\begin{eqnarray}
F^\alpha[h] &=& \sum_l F^{(bare)\alpha}_l(t,r) \,. \label{eq:bsf}
\end{eqnarray}
We note that the harmonic decomposition is ill-defined 
on a $2$-sphere of $(t,r)=(t_0,r_0)$ 
because the bare self-force field diverges, 
as a result, the mode decomposition is not continuously defined 
across this 2-sphere as 
$\lim_{t\to t_0,r\to r_0+0}F^{(bare)\alpha}_l \,\not =\, 
\lim_{t\to t_0,r\to r_0-0}F^{(bare)\alpha}_l $. 

As for the singular self-force field to be subtracted, 
we only have its local expression by the Hadamard expansion, 
and it is not defined on the entire 2-sphere $(\theta,\phi)$, 
hence, an exact harmonic decomposition cannot be defined. 
However, the regularization calculation (\ref{eq:sf}) 
depends only on the local behavior of the fields around the orbit, 
and one can freely add a finite term to the singular self-force field 
if it vanishes at the coincidence limit $x^\alpha \to z^\alpha$. 
We consider the local coordinate expansion of the singular self-force field. 
We suppose that $x^\alpha-z^\alpha$ is sufficiently small 
and expand the singular self-force field 
by the power of $x^\alpha-z^\alpha$ as 
\begin{eqnarray}
F^\alpha[h^{sing}] &=& \sum_{n,n_\alpha}
f^\alpha_{(n,n_\alpha)}{1\over \epsilon^{n/2}}(t-t_0)^{n_t}(r-r_0)^{n_r}
(\theta-\theta_0)^{n_\theta}(\phi-\phi_0)^{n_\phi} \label{eq:sing} \,, \\ 
\epsilon &=& {1\over 2}g_{\alpha\beta}(z)
(x^\alpha-z^\alpha)(x^\beta-z^\beta) \,, 
\end{eqnarray}
where $g_{\alpha\beta}(z)$ is
the background metric evaluated at the orbital location. 
For the regularization calculation, it is sufficient 
to decompose the non-vanishing terms of (\ref{eq:sing}), 
i.e. only the terms of $n_t+n_r+n_\theta+n_\phi-n\leq 0$. 
Necessary formulae are developed in Ref.\cite{self}, 
and we have the mode decomposition 
of the singular self-force field schematically 
with vanishing ambiguous terms as 
\begin{eqnarray}
F^\alpha[h^{sing}] &\deq& \sum_l F^{(sing')\alpha}_l(t,r) \,. \label{eq:ssf}
\end{eqnarray}
(Because of vanishing ambiguous terms, LHS and RHS are not exactly equal.)  
Again, the expansion coefficients are not continuous 
at the coincidence limit$(t,r) \to (t_0,r_0)$. 

We subtract the singular self-force field from the bare self-force field. 
Because the singular behavior of these fields cancel each other, 
the sum of the harmonics becomes finite 
even after taking the coincidence limit, 
thus, the self-force is derived by 
\begin{eqnarray}
F^\alpha &=& \sum_l 
\{F^{(bare)\alpha}_l(t_0,r_0)-F^{(sing')\alpha}_l(t_0,r_0)\} \,.
\end{eqnarray}
This shows that all we need in the end are 
the expansion coefficients at the coincidence limit $(t,r)\to(t_0,r_0\pm 0)$, 
where the limit must be taken consistently.

%%%%%%%%%%%%%%%%%%%%%%%%%%%%%%%%%%%%%%%%%%%%%%%%%%%%%%%%%%%
\subsection{Radiation Reaction Formula} \label{ssec:rr}
%%%%%%%%%%%%%%%%%%%%%%%%%%%%%%%%%%%%%%%%%%%%%%%%%%%%%%%%%%%

The primary question to motivate the radiation reaction formula
is a relation between the self-force and the balance formula.
As we argue the self-force calculation in Subsec.\ref{ssec:sfc},
a complicated regularization calculation is necessary 
to derive the self-force, 
while the balance formula is simply formulated without regularization.
We also have a question which part of the self-force is
described in the balance formula.
We find the relation of the self-force and the balance formula
in Ref.\cite{RR}, and review some basic argument here.
We also refer Ref.\cite{gal}
for a hint to answer these questions, 
where the energy and angular momentum loss of the balance formula 
are derived from the radiative part of the self-force 
by taking an infinite time average. 

For preparation to discuss the radiation reaction formula,
we first argue some symmetry property of
a Kerr geometry and geodesics.
Using constants of motion ${\cal E}^a,\,a=\{E,L_z,C\}$,
the geodesic equation in the Boyer-Lindquist coordinates becomes
\begin{eqnarray}
\left({dr \over d\lambda}\right)^2 \,=\, R({\cal E}^a;r) \,, && \quad
\left({d\theta\over d\lambda}\right)^2 \,=\,
\Theta({\cal E}^a;\theta) \,, \label{eq:geo1} \\
{dt \over d\lambda} \,=\,
T_r({\cal E}^a;r)+T_\theta({\cal E}^a;\theta) \,, && \quad
{d\phi \over d\lambda} \,=\, 
\Phi_r({\cal E}^a,r)+\Phi_\theta({\cal E}^a;\theta)
\,. \label{eq:geo2} 
\end{eqnarray}
As we are interested in the inspiral stage of the binary,
we only consider the case that $r$-motion is bounded,
then, the $r$- and $\theta$-motions become periodic\cite{cha}
and we can expand them by discrete Fourier series as
\begin{eqnarray}
z^b &=& \sum_n Z^{b(n)} \exp[in\chi_b] \,, \quad
\chi_b \,=\, 2\pi\tilde\Omega_b(\lambda+\lambda^b)
\,, \label{eq:geo_r}
\end{eqnarray}
where $b=\{r,\theta\}$ and $r=z^r,\,\theta=z^\theta$,
and, $Z^{b(n)}$ and $\tilde\Omega_b$ are functions of ${\cal E}^a$.
Here we have two integral constants $\lambda^b$,
but, since we can freely choose the zero point of $\lambda$,
only $\lambda^r-\lambda^\theta$ can specify the geodesic.
Because of these periodicities,
one can freely add $1/\tilde\Omega_b$ to $\lambda_b$.
Suppose that the ratio of $\tilde\Omega_r$ to $\tilde\Omega_\theta$
is irrational,
one can set $\lambda^r-\lambda^\theta$ as small as possible, 
then, by an appropriate choice of the zero point of $\lambda$,
one can have $\lambda^r=\lambda^\theta=0$.
$t$- and $\phi$-motions are simply integrated as
\begin{eqnarray}
z^c &=& \kappa_c+\sum_{b,n}Z_b^{c(n)}\exp[in\chi_b] \,, \quad
\kappa_c \,=\, <\dot Z^c>\lambda+C^c
\,, \label{eq:geo_the}
\end{eqnarray}
where $c=\{t,\phi\}$ and $t=z^t,\,\phi=z^\phi$,
and, $<\dot Z^c>$ and $Z_b^{c(n)}$ are function of ${\cal E}^a$.
Here we have two integral constants $C^c$
which specify the geodesic.
Below, we call ${\cal E}^a$ by primary constants, 
and, $\lambda^b$ and $C^c$ by secondary constants of geodesics.
The Kerr geometry has the $t$ and $\phi$-translation symmetries
and, by these properties, one can set $C^c=0$.

We find an interesting symmetry of the Kerr geometry.
The metric is invariant under
\begin{eqnarray}
t \to -t ,\, r \to r ,\, \theta \to \theta ,\, \phi \to -\phi
\,. \label{eq:gps}
\end{eqnarray}
By this symmetry, the orbital parameter and
the primary and secondary constants of a geodesic transform as
\begin{eqnarray}
\lambda \to -\lambda ,\, {\cal E}^a \to {\cal E}^a ,\,
\lambda^b \to -\lambda^b ,\, C^c \to -C^c \,.
\end{eqnarray}
Since the secondary constants can be set zero,
the geodesic is invariant under this symmetry transformation
and we call this by the geodesic preserving symmetry.

We next discuss the self-force.
Because the self-force is gauge dependent,
one has to specify a gauge condition for this discussion.
Here we restrict a class of gauge conditions 
such that the metric perturbation is derived as 
\begin{eqnarray}
h_{\mu\nu}(x) &=& \int dx' 
G_{\mu\nu\,\mu'\nu'}(t-t',\phi-\phi';r,r',\theta,\theta') 
T^{\mu'\nu'}(x') \,, \label{eq:prg}
\end{eqnarray}
where the Green function has an appropriate fall-off condition
at $|t-t'|,|r-r'|\to\infty$.
We call this by the physically reasonable class of gauge conditions 
because the Green function is invariant by 
the $t$ and $\phi$-translation symmetry of the Kerr geometry, 
and because one can naturally read out 
a spectral information of asymptotic gravitational waves 
in this class of gauge conditions.
Instead of the coordinate components of the self-force,
we consider the evolution of the "primary constants" by the self-force. 
The primary constants of a geodesic are defined 
by the Killing vectors $\eta^{E/L}_\alpha$ 
and tensor $\eta^C_{\alpha\beta}$ as 
$E=\eta^E_\alpha v^\alpha,L=\eta^L_\alpha v^\alpha,
C=\eta^C_{\alpha\beta} v^\alpha v^\beta/2$, 
and we consider the self-force acting on the "primary constants" 
$\tilde F^a = d{\cal E}^a/d\lambda$ derived as 
\begin{eqnarray}
\tilde F^E \,=\,
\eta^E_\alpha {D\over d\lambda}v^\alpha \,, \quad
\tilde F^L \,=\,
\eta^L_\alpha {D\over d\lambda}v^\alpha \,, \quad
\tilde F^C \,=\,
\eta^C_{\alpha\beta} v^\alpha {D\over d\lambda}v^\beta 
\,. \label{eq:sfc}
\end{eqnarray}
We note that (\ref{eq:geo1}) and (\ref{eq:geo2}) rely 
only on the definitions of ${\cal E}^a$ 
whether they are constants or not, 
thus, (\ref{eq:geo1}), (\ref{eq:geo2}) and (\ref{eq:sfc}) 
describe the orbital evolution by the self-force. 
Because the self-force is induced by a geodesic to the leading order,
it is a function of 
the primary and secondary constants and the orbital parameters. 
We also find that the self-force does not depend on $C^c$ 
because of the $t$ and $\phi$-translation symmetries,
thus, we have 
$\tilde F^a = \tilde F^a({\cal E}^a,\lambda^b,C^c;\lambda)$. 
Under the gauge condition defined by (\ref{eq:prg}), 
we find the self-force could be expanded by discrete Fourier series as 
\begin{eqnarray}
\tilde F^a &=& \sum_{n_r,n_\theta}
\dot{\cal E}^{a(n_r,n_\theta)}\exp[in_r\chi_r+in_\theta\chi_\theta]
\,. \label{eq:sfx}
\end{eqnarray}

We consider to apply 
the geodesic preserving symmetry transformation (\ref{eq:gps}) 
to the self-force.
Because the transformation changes the direction of time,
it also changes the boundary condition to derive the self-force.
We derive the key identity, 
\begin{eqnarray}
\tilde F^{(ret)a}({\cal E}^a,\lambda^b;\lambda) &=&
-\tilde F^{(adv)a}({\cal E}^a,-\lambda^b;-\lambda) \,,
\end{eqnarray}
where we note the retarded boundary condition 
and the advanced boundary condition 
by $(ret)$ and $(adv)$ respectively 
to indicate the boundary condition for the self-force. 
Using the general form of the self-force (\ref{eq:sfx}), we have
$\dot{\cal E}^{(ret)a(n_r,n_\theta)} =
-\dot{\cal E}^{(adv)a(-n_r,-n_\theta)}$.
Suppose we calculate the self-force 
by the radiative (half-retarded-minus-half-advanced) Green function,
we would have
\begin{eqnarray}
{1\over 2}\left(\dot{\cal E}^{(ret)a(n_r,n_\theta)}
+\dot{\cal E}^{(ret)a(-n_r,-n_\theta)}\right) &=&
\dot{\cal E}^{(rad)a(n_r,n_\theta)} \,,
\end{eqnarray}
where we note the radiative boundary condition by $(rad)$ 
to indicate the boundary condition for the self-force. 
This shows that we obtain the half of the self-force
by using the radiative metric perturbation.
The balance formula by Ref.\cite{PT} and Ref.\cite{gal}
discusses the infinite time averaged loss
of the energy and angular momentum, corresponding to
$\dot{\cal E}^{(ret)a(0,0)}=\dot{\cal E}^{(rad)a(0,0)}$,
but, the present derivation also includes the Carter constant.

This result suggests why the balance formula was derived
without a complicated regularization calculation
as was required in the self-force calculation.
Because the radiative Green function is defined as
${\bf G}^{rad}=({\bf G}^{ret}-{\bf G}^{adv})/2$\cite{gal},
the singular parts of the retarded Green function
and the advanced Green function cancel each other,
thus, the regularization calculation is naturally done
in the balance formula\cite{PT}
and the radiation reaction formula\cite{RR}.

Using the general form of the self-force (\ref{eq:sfx}),
one can derive a general property of the orbit
with the radiation reaction effect 
by integrating (\ref{eq:geo1}), (\ref{eq:geo2}) and (\ref{eq:sfc}).
Here, it is important to observe that
the self-force is derived by the linear metric perturbation. 
The divergence of the linearized Einstein tensor is known to 
vanish algebraically, 
thus, the stress-energy tensor of its source term conserves 
in the background. 
For this reason, the use of a linear metric perturbation 
as an approximation of the geometry 
is valid only when the deviation from a geodesic is small enough, 
and, in Ref.\cite{RR}, we consider the orbit 
by a perturbation from a geodesic. 
By this perturbative analysis of the orbit,
we find the evolution of the "primary constants" 
and the "secondary constants" by the gauge condition (\ref{eq:prg}) 
can be written as
\begin{eqnarray}
{\cal E}^a &=&
{\cal E}^a_0+<\dot {\cal E}^a> \lambda
+\sum_{n_r,n_\theta}
{\cal E}^{a(n_r,n_\theta)}\exp[in_r\chi_r+in_\theta\chi_\theta]
\,, \label{eq:ee} \\
\lambda^b &=&
\lambda^b_0 +<\ddot\lambda^b> \lambda^2/2
+\sum_{n_r,n_\theta}
(\dot\lambda^{b(n_r,n_\theta)}\lambda+\lambda^{b(n_r,n_\theta)})
\exp[in_r\chi_r+in_\theta\chi_\theta]
\,, \label{eq:ll} \\
C^c &=&
C^c_0 +<\ddot C^c> \lambda^2/2
+\sum_{n_r,n_\theta}
(\dot C^{c(n_r,n_\theta)}\lambda+C^{c(n_r,n_\theta)})
\exp[in_r\chi_r+in_\theta\chi_\theta]
\,, \label{eq:cc}
\end{eqnarray}
where ${\cal E}^a_0,\lambda^b_0,C^c_0$ are
the initial values of the "constants".
It is notable that
$<\dot {\cal E}^a>$, $<\ddot\lambda^b>$ and $<\ddot C^c>$
determine the dominant contribution of
(\ref{eq:ee}), (\ref{eq:ll}) and (\ref{eq:cc}), 
which are derived only by the radiative metric perturbation.
We also prove that these quantities are gauge invariant 
in the physically reasonable class of gauge conditions. 

Because the radiative Green function is a homogeneous Green function, 
there is a convenient method to calculate it 
even in a Kerr black hole\cite{chrz}
in the so-called radiation gauge conditions.
A necessary calculation here is essentially 
a minor extension of that used for the balance formula, 
and we consider that 
a numerical method is already established\cite{pp}.

%%%%%%%%%%%%%%%%%%%%%%%%%%%%%%%%%%%%%%%%%%%%%%%%%%%%%%%%%%%
\subsection{Self-Force Calculation vs Radiation Reaction Formula}
%%%%%%%%%%%%%%%%%%%%%%%%%%%%%%%%%%%%%%%%%%%%%%%%%%%%%%%%%%%

Here we try to compare these two approaches to calculate the orbit.
There could be two issues for comparison;\\
{\bf 1)}
{\it Theoretical issue; whether the resulting orbit
is physically acceptable or not} \\
{\bf 2)}
{\it Practical issue; whether the approach is practically available
for generating templates for LISA project} \\

At this moment, we may say that
both the self-force calculation and the radiation reaction formula
still have the theoretical issue.

Gravitational radiation reaction is a physically real object.
Because the momentum flux carried by gravitational waves is
well defined at the asymptotic flat region of the background spacetime,
it is reasonable to expect 
the orbital evolution must be consistent with this effect.
There is a wide-spread belief that 
the self-force includes the effect of gravitational radiation reaction. 
Although this looks a reasonable conclusion, 
we find that the relation 
between the gravitational self-force and gravitational radiation reaction 
in a usual metric perturbation scheme 
is not trivial because of the gauge freedom. 
As we shall argue in Sec.\ref{sec:sf?}, 
the self-force totally depends on a gauge condition
in the whole time interval 
where a usual metric perturbation scheme is valid,
thus, the gravitational self-force does not necessarily 
describe the effect of gravitational radiation reaction to the orbit 
as we have expected. 
In Sec.\ref{sec:ad}, we shall discuss that 
this contradicting problem appears 
because the usual metric perturbation scheme is not appropriate 
to consider radiation reaction to the orbital evolution, 
and we shall propose 
an adiabatic approximation of the metric perturbation 
where the physically reasonable class of gauge conditions is adopted. 
Unless the self-force calculation is done 
in the physically reasonable class of gauge conditions, 
we consider it necessary to argue that 
the self-force calculation in a gauge condition 
describe an orbital evolution by radiation reaction 
in an appropriate manner. 

On the other hand,
the radiation reaction formula is derived consistently 
with the balance formula 
by using the physically reasonable class of gauge conditions. 
however, the radiation reaction formula 
gives us only the infinite time averaged part of the self-force 
acting on the "primary constants". 
Here we have a question. 
A general form of the self-force (\ref{eq:sfx}) 
shows that the self-force has an oscillating component 
other than the infinite time averaged component. 
It is not clear how this oscillating component 
contributes to the orbital evolution. 
(\ref{eq:ee}), (\ref{eq:ll}) and (\ref{eq:cc}) 
show that the infinite time averaged component 
dominantly determines the orbital evolution, 
however, this analysis is based 
on a usual metric perturbation scheme. 
As we shall argue in Sec.\ref{sec:sf?}, 
the usual metric perturbation is not appropriate 
to consider the orbital evolution, 
and we shall propose 
an adiabatic approximation of the metric perturbation 
in Sec.\ref{sec:ad}. 
Then the question remains on a possibility 
that the oscillating component might be important 
to consider the orbital evolution. 

As we shall discuss in Sec.\ref{sec:ad},
we consider the radiation reaction formalism 
under the adiabatic approximation of the metric perturbation. 
We find that the infinite time averaged part of the self-force
actually determines the orbit during the time scale 
where the adiabatic approximation of the metric perturbation holds. 
We also find that the oscillating part of the self-force 
can be eliminated by a special choice of a gauge condition 
in Ref.\cite{RRx}.

As for the practical issue, 
still the self-force calculation has a lot to investigate. 
There are a number of studies 
of a linear metric perturbation induced by a point particle\cite{pp}. 
however, most of these studies calculate gravitational waves at infinity 
where both a numerical and a semi-analytic methods are established. 
On the other hand, the self-force calculation needs 
a mode decomposition of a linear metric perturbation along the orbit 
in order to derive the bare self-force field (\ref{eq:bsf}) 
and a reliable technique for this calculation 
has not been demonstrated in general. 

If the background is a Schwarzschild black hole, 
one can use the Zerilli-Regge-Wheeler formalism\cite{ZRW} 
to derive a mode decomposition of the bare self-force field 
for a general orbit in principle, 
and we have a result for a circular orbit or a radial orbit, 
not for a general orbit. 
But, when the background is a Kerr black hole, 
a method to calculate a metric perturbation 
is just proposed\cite{amos}, 
and the self-force calculation with this idea is not yet formulated. 
A more crucial problem here is that 
it is not clear whether 
the self-force calculation using this metric perturbation 
would describe the effect of gravitational radiation reaction 
in a physically reasonable manner 
because the gauge condition along the orbit is not well understood. 

We also comment that 
the self-force calculation is a point splitting regularization, 
and that the singular self-force field might not 
be optimally subtracted from the bare self-force field. 
As a result, a numerical convergence of the self-force calculation 
might not be fast in general.

On the other hand, 
the radiation reaction formula can be seen as 
a non-trivial extension of the balance formula. 
We only need a radiative metric perturbation 
and a convenient method for this calculation is known\cite{chrz}. 
Various numerical codes of the balance formula
are already made successfully\cite{pp} 
for either a circular orbit or an equatorial orbit 
around a Kerr black hole, 
and we are coming to have a result for a general orbit, 
therefore, we consider this approach is practically promising. 
There is also known a semi-analytic technique
which may substantially increase the efficiency of the code\cite{mano}. 
We also comment that we have 
the most optimal convergence in this approach.

%%%%%%%%%%%%%%%%%%%%%%%%%%%%%%%%%%%%%%%%%%%%%%%%%%%%%%%%%%%
%%%%%%%%%%%%%%%%%%%%%%%%%%%%%%%%%%%%%%%%%%%%%%%%%%%%%%%%%%%
\section{Self-force? a myth?} \label{sec:sf?}
%%%%%%%%%%%%%%%%%%%%%%%%%%%%%%%%%%%%%%%%%%%%%%%%%%%%%%%%%%%

So far, we frequently noted that 
the self-force does not necessarily include 
the effect of gravitational radiation reaction, 
which is obviously against 
our motivation to study the self-force problem.
Because it has not been seriously considered 
whether the self-force includes 
the effect of gravitational radiation reaction,
we shall devote this section to discuss this problem. 

Gravitational radiation reaction is physically real
since we can define a momentum flux of gravitational waves 
at the asymptotic flat region of the background metric, 
however, we find that the self-force is entirely gauge dependent 
and that it could vanish by a special gauge choice along the orbit. 
It is well-known that the self-force could vanish 
at an orbital point by a gauge choice, 
but we considered this might not be a problem 
because the self-force would have non-vanishing components 
of gravitational radiation reaction 
by taking an infinite time average of the self-force. 
In Subsec.\ref{ssec:rr}, we discuss that this is the case 
in the physically reasonable class of gauge conditions 
and that non-vanishing components are proven to be 
gauge invariant in this gauge class. 
However, a usual metric perturbation scheme 
can describe a gravitational evolution of the system 
only for a finite time interval, 
and, in this whole time interval, 
the self-force could even vanish by a gauge choice along the orbit 
as we show in Subsec.\ref{ssec:per}. 
This suggests that the relation between 
the self-force and gravitational radiation reaction 
is entirely gauge dependent, 
and is not trivial as we expected before. 
Then we have questions; 
how the self-force is related with gravitational radiation reaction, 
and what gauge conditions we should take to calculate a self-force. 
We consider these questions in Subsec.\ref{ssec:eng} 

%%%%%%%%%%%%%%%%%%%%%%%%%%%%%%%%%%%%%%%%%%%%%%%%%%%%%%%%%%%
\subsection{Metric Perturbation} \label{ssec:per}
%%%%%%%%%%%%%%%%%%%%%%%%%%%%%%%%%%%%%%%%%%%%%%%%%%%%%%%%%%%

We suppose that 
a regularization calculation for a self-force 
is formally possible by a matched asymptotic expansion 
as we discuss in Subsec.\ref{ssec:mae}, 
and consider only a metric of the far-zone expansion. 
The metric of the far-zone expansion is a sum 
of a regular vacuum background metric and its metric perturbation 
induced by a point particle moving on the background metric. 
We suppose the point particle has an appropriate structure\cite{Di} 
so that the metric of the far-zone expansion 
consistently matches a metric of the near-zone expansion. 

Suppose we use a usual metric perturbation scheme 
for the calculation of the metric of the far-zone expansion. 
We expand the metric and the stress-energy tensor of the point particle 
with a small parameter $m/L$ as 
\begin{eqnarray}
g_{\mu\nu} &=& g^{(bg)}_{\mu\nu}
+(m/L) h^{(1)}_{\mu\nu}+(m/L)^2 h^{(2)}_{\mu\nu}+\cdots \,, \\
T^{\mu\nu} &=& (m/L) T^{(1)\mu\nu}
+(m/L)^2 T^{(2)\mu\nu}+\cdots \,,
\end{eqnarray}
where $g^{(bg)}_{\mu\nu}$ is the vacuum background metric. 
For a valid perturbation, we have 
\begin{eqnarray}
&& O(1) > (m/L) h^{(1)}_{\mu\nu} > 
(m/L)^2 h^{(2)}_{\mu\nu} > \cdots \,, \\
&& O(1) > (m/L) T^{(1)\mu\nu} > 
(m/L)^2 T^{(2)\mu\nu} > \cdots \,. 
\end{eqnarray}
One can expand the Einstein equation with $m/L$, 
and we schematically have 
\begin{eqnarray}
&& G^{(1)\mu\nu}[h^{(1)}] 
\,=\, T^{(1)\mu\nu} \,, \label{eq:ein1} \\
&& G^{(1)\mu\nu}[h^{(2)}]
+G^{(2)\mu\nu}[h^{(1)},h^{(1)}]
\,=\, T^{(2)\mu\nu} \,, \label{eq:ein2} \\
&& \cdots \,, \nonumber
\end{eqnarray}
where $G^{(1)\mu\nu}[h]$ and $G^{(2)\mu\nu}[h]$ 
are the terms linear and quadratic in $h_{\mu\nu}$ 
of the Einstein tensor $G^{\mu\nu}[g+h]$, respectively. 

As we discuss in Subsec.\ref{ssec:mae}, 
$T^{(1)\mu\nu}$ is a usual stress-energy tensor of a monopole particle. 
It is important to note that 
the linearized Einstein equation is algebraically divergence free 
with respect to the background metric as 
$G^{(1)\mu\nu}{}_{;\nu}=T^{(1)\mu\nu}{}_{;\nu}=0$, 
thus, in this metric perturbation scheme, 
the particle as a source of $h^{(1)}_{\mu\nu}$ 
must move along a geodesic 
in the background metric $g^{(bg)}_{\mu\nu}$ 
for a consistent solution of (\ref{eq:ein1}). 
An explicit form of $T^{(2)\mu\nu}$ may be derived 
by doing a matched asymptotic expansion of metrices 
of the far-zone expansion and the near-zone expansion to this order. 
But, at least, one may consider that it has a term of 
the monopole particle deviating from a geodesic 
because one can derive the MiSaTaQuWa self-force 
from LHS of (\ref{eq:ein2}) by a mass renormalization\cite{sf}. 
Thus, for the expansion of the stress-energy tensor, 
we must also consider the expansion of the orbit as 
\begin{eqnarray}
z^\mu(\tau) &=& z^{(bg)\mu}(\tau)
+(m/L) z^{(1)\mu}(\tau)+\cdots \,,\label{eq:oe}
\end{eqnarray}
where $\tau$ is defined to be a proper time in the background metric 
and $z^{(bg)\mu}$ is a geodesic of the background metric. 
For a valid perturbation, we have 
\begin{eqnarray}
&& O(1) > (m/L) z^{(1)\mu} > \cdots \,. \label{eq:ov} 
\end{eqnarray}
Intuitively speaking, the orbit deviates from a geodesic 
by gravitational radiation reaction, 
and eventually the condition (\ref{eq:ov}) would be violated, 
then the metric perturbation in this perturbation scheme 
would fail to approximate the system. 

We consider a gauge transformation in the time scale 
where the metric perturbation still approximates the system. 
The gauge transformation is defined by a small coordinate transformation 
$x^\mu \to \bar x^\mu = x^\mu+(m/L)\xi^\mu(x)$ 
and the orbit transforms as 
$ z^\mu(\tau) \to \bar z^\mu(\tau)
= z^\mu(\tau')+(m/L)\xi^\mu(z(\tau'))$,
where $\tau'$ is related with the orbital parameter $\tau$ 
so that $\tau$ remains a proper time 
of the new orbit $\bar z^\mu(\tau)$ in the background metric. 
Applying this to 
the perturbation expansion of the orbit (\ref{eq:oe}), 
we find the gauge transformation of $(m/L)z^{(1)\mu}$ as 
\begin{eqnarray}
(m/L)z^{(1)\mu}(\tau) \,\to\, (m/L)\bar z^{(1)\mu}(\tau) 
\,=\, (m/L)z^{(1)\mu}(\tau)
+v^{(bg)\mu} (\tau'-\tau)+(m/L)\xi^\mu(z^{(bg)}(\tau)) 
\,, \label{eq:oo} 
\end{eqnarray}
where $v^{(bg)\mu}=dz^{(bg)\mu}/d\tau$. 
We find that we could eliminate $(m/L) z^{(1)\mu}(\tau)$ 
if the gauge transformation along the orbit satisfies 
\begin{eqnarray}
(m/L)\xi^\mu(z^{(bg)}(\tau)) &=& 
-(m/L)z^{(1)\mu}(\tau)+v^\mu \delta\tau \,, 
\end{eqnarray}
with an arbitrary small function $\delta\tau \sim O(m/L)$. 
There exists a gauge transformation 
which satisfies this condition for the whole time interval 
where the metric perturbation scheme is valid 
because it is smaller than $O(1)$, 
and the orbit becomes a geodesic of the background metric 
in this gauge condition. 

Because the self-force is an self-acceleration 
to deviate the orbit from a geodesic, 
$(m/L)z^{(1)\mu}(\tau)=0$ in this gauge condition means 
that the self-force entirely vanishes for the whole time interval 
where the metric perturbation scheme is valid. 
This extreme example suggests that 
the self-force is a totally gauge dependent object 
in the usual metric perturbation scheme, 
and, even a time average of the self-force 
over a longest time scale where the metric perturbation is valid 
is gauge dependent in general. 

The essential reason why the self-force could not include
the effect of gravitational radiation reaction is
that this perturbation scheme only allows a small deviation
from a geodesic as in (\ref{eq:ov}).
Because of this, one could always eliminate the deviation
by a gauge transformation as in (\ref{eq:oo}).
We note that this problem may not be solved 
by calculating a non-linear metric perturbation 
because the problem comes from the metric perturbation scheme. 
We consider a solution of this problem is 
to modify the metric perturbation scheme 
so that one can describe 
a non-perturbative orbital deviation from a geodesic. 
If the orbit can deviate from a geodesic non-perturbatively, 
one cannot eliminate it by a gauge transformation. 

%%%%%%%%%%%%%%%%%%%%%%%%%%%%%%%%%%%%%%%%%%%%%%%%%%%%%%%%%%%
\subsection{Energetics of the Orbit and Radiation} \label{ssec:eng}
%%%%%%%%%%%%%%%%%%%%%%%%%%%%%%%%%%%%%%%%%%%%%%%%%%%%%%%%%%%

A hint to understand why such an unexpected thing happens 
can be seen in Ref.\cite{RR}. 
In Sec.III of Ref.\cite{RR}, 
we argue that the orbital energy does not decrease monotonically 
by the emission of gravitational waves. 
We consider that gravitational radiation has its own energy 
and that the self-force describes the interaction 
between the orbital energy and the radiation energy. 
In fact, one can define an effective stress-energy tensor 
of the orbit and gravitational radiation\cite{sf}. 
As in the previous subsection, 
we only consider the metric of the far-zone expansion, 
and we suppose that 
the metric is a sum of the vacuum background metric 
and its perturbation induced by a point particle 
with an appropriate internal structure\cite{Di} 
as $g_{\mu\nu}= g^{(bg)}_{\mu\nu}+h_{\mu\nu}$. 

The Einstein equation can formally written as 
$G^{(1)}_{\mu\nu}[h]+G^{(2+)}_{\mu\nu}[h]=T_{\mu\nu}$, 
where $G^{(1)}_{\mu\nu}$ and $G^{(2+)}_{\mu\nu}$ 
are the linear terms and the rest of the Einstein tensor 
with respect to $h_{\mu\nu}$, 
and $T_{\mu\nu}$ is the stress-energy tensor of the point particle. 
Because $G^{(1)}_{\mu\nu}$ is algebraically divergence free, 
one can define a conserved stress-energy tensor 
in the background metric by 
\begin{eqnarray}
{\cal T}^{\mu\nu} &=& T^{\mu\nu}-G^{(2+)\mu\nu}[h]
\,. \label{eq:einx} 
\end{eqnarray}
We consider that 
the first term and the second term of (\ref{eq:einx}) represent 
the stress-energy tensors of the particle and gravitational radiation 
respectively, 
and each of these are not conserved in the background by themselves. 
We suppose that 
the background metric $g^{(bg)}_{\mu\nu}$ is a Kerr black hole 
and has a timelike Killing vector $\xi_\mu$, 
then one can define a total, orbital and radiation energy 
in the background metric as
\begin{eqnarray}
E^{(tot)} \,=\, E^{(orb)}+E^{(rad)} \,,\quad 
E^{(orb)} \,=\, -\int d\Sigma_\mu \xi_\nu T^{\mu\nu} \,, \quad 
E^{(rad)} \,=\, \int d\Sigma_\mu \xi_\nu G^{(2+)\mu\nu} 
\,, \label{eq:ene} 
\end{eqnarray}
where the surface integration is taken 
over a spacelike hypersurface 
bounded by the future horizon and the future null infinity 
of the background black hole metric. 
We note that 
the radiation energy $E^{(rad)}$ is not necessarily positive. 

By integrating (\ref{eq:einx}) 
over a small world tube surface around the orbit, 
one can derive the MiSaTaQuWa self-force\cite{sf}, 
which shows that the self-force describes the interaction 
between the orbital energy and the gravitational radiation energy. 
A gauge condition for the self-force could be 
interpreted as a small arbitrariness 
in defining the orbital energy and the radiation energy 
in (\ref{eq:ene})\cite{RRx}. 
In order to see this, 
we apply the metric perturbation scheme of Subsec.\ref{ssec:per} 
to (\ref{eq:ene}), and we have 
\begin{eqnarray}
E^{(orb)} &=& (m/L)E^{(orb)(1)}+(m/L)^2E^{(orb)(2)}+\cdots \,, 
\nonumber \\ 
(m/L)E^{(orb)(1)} &=& 
-\int d\Sigma_\mu \xi_\nu (m/L)T^{(1)\mu\nu} \,, \quad 
(m/L)^2E^{(orb)(2)} \,=\, 
-\int d\Sigma_\mu \xi_\nu (m/L)^2T^{(2)\mu\nu} \,, \quad \cdots \\ 
E^{(rad)} &=& (m/L)^2E^{(orb)(2)}+\cdots \,, \quad  
(m/L)^2E^{(rad)(2)} \,=\, 
\int d\Sigma_\mu \xi_\nu (m/L)^2 G^{(2)\mu\nu}[h^{(1)},h^{(1)}] 
\,, \quad \cdots \,, \label{eq:enex} 
\end{eqnarray}
In the usual perturbation scheme, 
$(m/L)T^{(1)\mu\nu}$ conserves by itself, 
as a result, $E^{(orb)(1)}$ becomes a constant of motion. 
By a gauge transformation around the orbit, one can show 
that the orbital energy and the radiation energy transform as 
\begin{eqnarray}
(m/L)^2E^{(orb)(2)} \,\to\, (m/L)^2\bar E^{(orb)(2)} 
\,=\, (m/L)^2E^{(orb)(2)}+(m/L)^2\delta E \,, \\
(m/L)^2E^{(rad)(2)} \,\to\, (m/L)^2\bar E^{(rad)(2)} 
\,=\, (m/L)^2E^{(rad)(2)}-(m/L)^2\delta E \,, 
\end{eqnarray}
where $\delta E$ is arbitrary, 
and the orbital energy is entirely gauge dependent 
by itself as we discuss in the previous section. 
It is important to note that $E^{(tot)}$ is gauge invariant\footnote{
Because of this property, one can use this as a criteria 
for a last stable orbit of an extreme mass ratio binary.}\cite{RRx}. 

Although we do not yet propose a new metric perturbation scheme, 
we may extend the argument of the orbital energy and the radiation energy 
since they are defined in a non-perturbative manner as (\ref{eq:ene}). 
The balance formula is derived by gravitational waves 
at the future null infinity and the future horizon 
of the background black hole geometry, 
and it describes radiation reaction to the total energy 
rather than the orbital energy\cite{RRx}. 
In the usual metric perturbation scheme, 
one can decrease only the radiation energy by radiation reaction 
while keeping the orbital energy constant by the gauge freedom. 
Although this is mathematically allowed 
in the usual metric perturbation scheme, 
if the system continues to lose the total energy by radiation reaction, 
it will eventually be difficult to understand the orbit 
in a physically reasonable way. 
Since we keep the orbital energy constant, 
the radiation energy will inevitably decrease 
substantially by radiation reaction, 
as a result, the amplitude of the metric perturbation 
would be non-perturbatively large 
and one cannot use the coordinate system of the background metric 
as a reference to observe the orbit. 

In this case, only numerical relativity may be able to describe 
the evolution of the system, 
and, even if it is possible, it may be difficult 
to understand the system. 
Thus, for this technical advantage, we consider it reasonable 
to keep the amplitude of the metric perturbation small 
and we should decrease the orbital energy 
by radiation reaction rather than the radiation energy. 
We consider that this would be a reasonable least criteria 
for a convenient gauge condition to calculate the self-force. 
Although we do not have a simpler criteria 
for a convenient gauge condition, 
the physically reasonable class of gauge conditions (\ref{eq:prg}) 
satisfies this criteria. 
Because the self-force in this gauge class is consistent 
with the balance formula, 
only the orbital energy decreases by radiation reaction 
and the radiation energy is just oscillating, 
as a result, the metric perturbation would not 
grow non-perturbatively large. 

Although the argument of energetic shows us 
an importance of the self-force, this is just a qualitative argument. 
In the next section, we propose a new perturbation scheme 
and we formulate a method to calculate a gravitational evolution 
of an extreme mass ratio binary in this picture.

%%%%%%%%%%%%%%%%%%%%%%%%%%%%%%%%%%%%%%%%%%%%%%%%%%%%%%%%%%%
%%%%%%%%%%%%%%%%%%%%%%%%%%%%%%%%%%%%%%%%%%%%%%%%%%%%%%%%%%%
\section{Adiabatic extension} \label{sec:ad}
%%%%%%%%%%%%%%%%%%%%%%%%%%%%%%%%%%%%%%%%%%%%%%%%%%%%%%%%%%%

In this section, we review our recent progress
on the radiation reaction formalism in Ref.\cite{RRx}.
We discuss in Sec.\ref{sec:sf?} that the self-force
does not necessarily include
the effect of gravitational radiation reaction
because of the perturbation scheme.
In order to break this limitation,
we introduce an adiabatic approximation to this problem.
The adiabatic approximation is well-known
in classical mechanics,
however, the application to gauge field theory
such as a gravitational perturbation is not so common.

%%%%%%%%%%%%%%%%%%%%%%%%%%%%%%%%%%%%%%%%%%%%%%%%%%%%%%%%%%%
\subsection{Adiabatic Metric Perturbation}
%%%%%%%%%%%%%%%%%%%%%%%%%%%%%%%%%%%%%%%%%%%%%%%%%%%%%%%%%%%

From the result of the radiation reaction formula,
the orbital deviation becomes $O(1)$
when $\lambda \sim O((m/L)^{-1/2})$.
This time scale is called a dephasing time of a orbit.
The usual metric perturbation scheme 
is valid only within this time scale 
as we discuss in Sec.\ref{sec:sf?}.
Our purpose here is to extend this time scale of validity
by modifying the metric perturbation scheme.
For an explicit discussion, we use 
the physically reasonable class of gauge condition (\ref{eq:prg}). 

We denote the orbital constants of a geodesic by
$\gamma = \{{\cal E}^a,\lambda^b,C^c\}$.
Because a linear metric perturbation is induced by a geodesic,
it is a function of $\gamma$ as $h_{\mu\nu}(x)=h_{\mu\nu}(\gamma;x)$.
In the radiation reaction formula\cite{RR},
we use an adiabatic approximation to the orbit. 
In this approximation, 
we approximate the orbit by a geodesic at each instant,
and consider the evolution of the "orbital constants"
by the effect of gravitational radiation reaction,
thus, we may write the "orbital constants"
as functions of the orbital parameter as $\gamma(\lambda)$.
We consider to extend this idea to the metric perturbation.
We foliate the spacetime by spacelike hypersurfaces 
which intersect with the orbit. 
We define the foliation function $f(x)$ by the orbital parameter
at the intersection of the orbit and the surfaces as
$f(z(\lambda))=\lambda$.
We define an adiabatic linear metric perturbation
by using the linear metric perturbation
on the foliation surface induced by the geodesic $\gamma(f)$ as
\begin{eqnarray}
h^{ad}_{\mu\nu}(x) &=& h_{\mu\nu}(\gamma(f);x) 
\label{eq:adm} \,. 
\end{eqnarray}
Here we do not specify an explicit form of the foliation function, 
as a result, an adiabatic linear metric perturbation 
is not defined uniquely. 
We consider that we may need a constraint of the foliation function 
for an adiabatic non-linear metric perturbation, 
however, it does not change the result of the following discussion. 
We consider that the adiabatic linear metric perturbation is
a non-trivial extension of the linear metric perturbation.

Under the physically reasonable class of gauge conditions, 
a formal expression of the tensor Green function becomes 
\begin{eqnarray}
G_{\mu\nu\,\mu'\nu'}(x,x') &=& 
\sum_{\omega,m}g_{\mu\nu\,\mu'\nu'}^{(\omega,m)}(r,\theta;r',\theta')
\exp[-i\omega(t-t')+im(\phi-\phi')] \,. 
\end{eqnarray}
This gives a general form of the linear metric perturbation 
in this class of gauge conditions 
induced by a geodesic $\gamma$ as 
\begin{eqnarray}
h_{\mu\nu}(\gamma;x) &=& \sum_{\omega,m,n_r,n_\theta}
k^{(\omega,m,n_r,n_\theta)}(\gamma)
h_{\mu\nu}^{(\omega,m,n_r,n_\theta)}({\cal E}^a;r,\theta)
\exp[-i\omega t+im \phi] \,, \label{eq:metric} \\
k^{(\omega,m,n_r,n_\theta)}(\gamma)
&=& \int d\lambda \tilde k^{(\omega,m,n_r,n_\theta)}({\cal E}^a)
\exp[i\omega\kappa_t-im\kappa_\phi
-in_r\chi_r-in_\theta\chi_\theta] \,.
\end{eqnarray}
We construct the adiabatic linear metric perturbation (\ref{eq:adm}) 
with the general form of the linear metric perturbation (\ref{eq:metric}). 
When we operate the linearized Einstein operator, 
we have 
\begin{eqnarray}
G^{(1)}_{\mu\nu}[h^{ad}] &=& T_{\mu\nu}[\gamma(f)]
+\Lambda^{(1)}_{\mu\nu}[h^{ad}] \,, \label{eq:adme} 
\end{eqnarray}
where an extra term $\Lambda^{(1)}_{\mu\nu}$ appears 
because the adiabatic linear metric perturbation 
is not induced by a geodesic of the background. 
It is notable that $T^{\mu\nu}[\gamma(f)]$ 
is a stress-energy tensor of a point particle 
moving along the orbit 
with the effect of gravitational radiation reaction. 

The adiabatic metric linear perturbation solves 
the Einstein equation to an accuracy of $O(m/L)$ 
as long as $O(m/L)>\Lambda^{(1)}_{\mu\nu}$ holds. 
Since $\Lambda^{(1)}_{\mu\nu}$ is proportional to
the $\lambda$-derivatives of the orbital "constants" 
$\gamma(\lambda)$, 
the validity of the adiabatic metric perturbation 
depends on how the orbit evolves by a self-force. 
In the next section, we discuss the orbital evolution 
under the physically reasonable class of gauge conditions. 
We find that the adiabatic metric perturbation 
is a well approximated solution 
of the linearized Einstein equation 
in the radiation reaction time scale $O((m/L)^{-1})>\lambda$, 
hence, the time scale of validity is much longer 
than that of the usual metric perturbation scheme. 

%%%%%%%%%%%%%%%%%%%%%%%%%%%%%%%%%%%%%%%%%%%%%%%%%%%%%%%%%%%
\subsection{Adiabatic Evolution of the Orbit}
%%%%%%%%%%%%%%%%%%%%%%%%%%%%%%%%%%%%%%%%%%%%%%%%%%%%%%%%%%%

Different from the usual perturbation scheme 
one can not use the perturbation of the orbit any more, 
and we need to calculate the orbit in a non-perturbative manner. 
Because of a technicality of a non-perturbative calculation, 
we only summarize the result 
and we suggest the readers to refer Ref.\cite{RRx} for the details. 

The self-force defined by the adiabatic linear metric perturbation
is simply derived as 
\begin{eqnarray}
{d\over d\lambda}{\cal E}^a &=& \tilde F^a
({\cal E}^a(\lambda),\lambda^b(\lambda),C^c(\lambda);\lambda) 
+O((m/L)^2) \,, 
\end{eqnarray}
and we can use the result of 
the original radiation reaction formula\cite{RR} 
in the radiation reaction time scale $\lambda <O((m/L)^{-1})$. 
As for the "primary constants", 
one can still deal with the evolution by a perturbation, 
and we find that $\dot{\cal E}^{a(0,0)}$ dominantly determines 
the orbital evolution. 
The subdominant parts of the "primary constants" are $O(m/L)$ 
and we may not see these effect by gravitational waves. 
The evolution of the "secondary constants" becomes 
non-perturbative beyond the dephasing time, 
however, we find the perturbation results of 
(\ref{eq:ll}) and (\ref{eq:cc}) are qualitatively correct. 
The dominant part of their evolution is described 
only by ${\cal E}^{a(0,0)}$, 
and the subdominant part grows linearly in $\lambda$. 

In summary, we find 
the qualitative behavior of the orbital evolution as 
\begin{eqnarray}
{d\over d\lambda}{\cal E}^a \,=\, O(m/L) \,, \quad 
{d\over d\lambda}\lambda^b \,=\, O((m/L)\lambda) \,, \quad 
{d\over d\lambda}C^c \,=\, O((m/L)\lambda) \,. 
\end{eqnarray}
This gives us 
the qualitative estimate of (\ref{eq:adme}) 
and we find $\Lambda^{(1)}_{\mu\nu} = O((m/L)^2\lambda)$. 
Hence, the adiabatic linear metric perturbation is valid
in the radiation reaction time scale $O((m/L)^{-1})>\lambda$.

%%%%%%%%%%%%%%%%%%%%%%%%%%%%%%%%%%%%%%%%%%%%%%%%%%%%%%%%%%%
\subsection{Radiation Reaction Gauge}
%%%%%%%%%%%%%%%%%%%%%%%%%%%%%%%%%%%%%%%%%%%%%%%%%%%%%%%%%%%

The validity of the adiabatic linear metric perturbation
depends on a behavior of $\Lambda^{(1)}_{\mu\nu}$
in (\ref{eq:adme}).
Because $\Lambda^{(1)}_{\mu\nu}$ is gauge dependent,
we may consider to extend the validity
by using a remaining gauge freedom
in the physically reasonable class of gauge conditions.

Using the adiabatic linear metric perturbation,
we find that the "primary constants" still evolve perturbatively as
\begin{eqnarray}
{\cal E}^a &=& {\cal E}^a_0 +<\dot {\cal E}^a>\lambda
+\sum_{n_r,n_\theta}
{\cal E}^{a(n_r,n_\theta)}\exp[in_r\chi_r+in_\theta\chi_\theta]
\,. \label{eq:ree}
\end{eqnarray}
By a gauge transformation $x^\mu \to x^\mu+\xi^\mu$,
the "primary constants" transforms as
${\cal E}^a \to {\cal E}^a +\delta {\cal E}^a$ where
\begin{eqnarray}
\delta {\cal E}^{E/L} &=&
-\eta^{E/L}_\alpha v^\beta \xi^\alpha_{;\beta}
+\eta^{E/L}_{\alpha;\beta} v^\beta \xi^\alpha
\,, \label{eq:e_gau} \\
\delta {\cal E}^C &=&
-\eta^C_{\alpha\beta} v^\beta v^\gamma \xi^\alpha_{;\gamma}
\,. \label{eq:c_gau}
\end{eqnarray}
In Ref.\cite{RRx}, we find that,
by an appropriate gauge choice,
one can eliminate the oscillating part of (\ref{eq:ree}) as
\begin{eqnarray}
{\cal E}^a &=& ({\cal E}^a_0+\delta{\cal E}^a_0)
+<\dot {\cal E}^a>\lambda +O(\mu^2 t) \,.
\end{eqnarray}
Here, one has to introduce
a small shift of the initial values $\delta{\cal E}^a_0$
so that the gauge transformation behaves as $\xi = O(m/L)$.
In this gauge condition,
the self-force has only the radiative part as
\begin{eqnarray}
\tilde F^a({\cal E}^a(\lambda),\lambda^b(\lambda),C^c(\lambda);\lambda)
\,=\, <\dot {\cal E}^a>({\cal E}^a(\lambda)) \,.
\end{eqnarray}
For this reason, we call this by a radiation reaction gauge.

The self-force is expected to have a conservative part
as well as the radiative part,
however, in this gauge condition,
the conservative part is integrated out 
to be a renormalization of the initial values
${\cal E}^a_0 \to {\cal E}^a_0+\delta{\cal E}^a_0$.
We also note that a radiation reaction gauge condition
is applicable to an orbit of a spinning particle.
It is known that the orbit of the spinning test particle
deviates from a geodesic\cite{papa}
by a coupling of its spin and the curvature of the background metric.
This effect can also be renormalized to the initial values
in the radiation reaction gauge condition.

Using a post-Newtonian estimation\cite{PN},
the self-force can be estimated to be $O(m/L \times v^5)$,
where $v^2 \sim 0.1$ around a last stable orbit. 
This is smaller than that in a general gauge condition
of the physically reasonable class.
As a result, one can predict the orbit in the time scale
$\lambda < O((m/L v^5)^{-1})$,
which may correspond to several years 
for a promising target of LISA.

%%%%%%%%%%%%%%%%%%%%%%%%%%%%%%%%%%%%%%%%%%%%%%%%%%%%%%%%%%%
%%%%%%%%%%%%%%%%%%%%%%%%%%%%%%%%%%%%%%%%%%%%%%%%%%%%%%%%%%%
\section{Summary and Future prospect} \label{sec:nxt}
%%%%%%%%%%%%%%%%%%%%%%%%%%%%%%%%%%%%%%%%%%%%%%%%%%%%%%%%%%%

It is widely believed that
a self-force could describe an orbital evolution 
with the effect of gravitational radiation reaction.
For this reason, the so-called Capra community were trying
to develop a method to explicitly calculate the self-force
acting on a particle orbiting around a Kerr black hole,
and we are having a success 
in developing a regularization calculation method 
either by the self-force calculation\cite{self}
or by the radiation reaction formula\cite{RR}.
On the other hand, we come to have a concern on
what physical information the self-force actually carries 
because the self-force is gauge dependent.
For this concern, it was discussed that it might be necessary
to develop a non-linear metric perturbation
to grasp a physical meaning of the self-force.

A second order metric perturbation may be derived consistently 
only when we consider an orbit 
with the effect of gravitational radiation reaction, 
and it gives us a gauge invariant information 
by an asymptotic gravitational waveform, 
for example, by a modulation of a gravitational wave phase. 
We consider that we may see 
the radiation reaction effect from the waveform, 
however, it has nothing to do 
with the radiation reaction effect to the orbit 
because radiation reaction to the orbit is purely gauge dependent. 
We find that one can define a radiation energy, an orbital energy, 
and a total energy as a sum of the radiation and orbital energy. 
Radiation reaction we usually consider (say, by the balance formula) 
is a reaction acting on the total energy, 
but, the orbital motion is related to the orbital energy. 

In Sec.\ref{sec:sf?}, 
we discuss that the gauge freedom is interpreted 
as a small ambiguity in separating the total energy 
into the radiation energy and the orbital energy. 
This separation is entirely arbitrary 
in the usual metric perturbation scheme, 
however, we discuss that, 
without some constraint on this separation, 
we may have a difficulty in interpreting the orbit. 
When we continue the evolution 
beyond the usual metric perturbation scheme, 
the metric perturbation would grow non-perturbatively large 
and, as a result, the background metric cannot be used 
as an approximated reference to track the orbit. 
We argue that the physically reasonable class of gauge conditions 
is, at least, a reasonable gauge choice 
to avoid such a non-perturbative situation in Sec.\ref{sec:sf?}, 
and we show that an adiabatic extension of 
the linear metric perturbation is possible in Sec.\ref{sec:ad}. 

A possible question here is
whether there is a wider class of gauge conditions
which allows an adiabatic extension of a linear metric perturbation.
Especially, we introduce
the physically reasonable class of gauge conditions
because a linear metric perturbation is derived 
by a Fourier-Harmonic decomposition, 
and it is not clear whether an adiabatic extension 
of a linear metric perturbation by a time domain calculation 
is possible or not. 
If it is not possible, 
the self-force calculation by the time domain calculation 
may not predict an orbital evolution 
in the radiation reaction time scale. 

Different from the self-force calculation,
the radiation reaction formula conveniently calculates 
only an infinite time averaged part of the self-force. 
We find that this part is actually enough 
for the prediction of an orbit 
in the radiation reaction time scale.
We find the rest of the self-force only makes
a small change of the orbit,
which can be totally eliminated by a choice of the gauge.

A crucial problem to calculate 
gravitational waveforms for the LISA project is
whether the radiation reaction time scale is sufficiently longer
than an observation time of an astrophysically expected target.
If we calculate the self-force
in the physically reasonable class of gauge conditions,
the radiation reaction time scale would around several months in general. 
Since an observation time of LISA would be around several years, 
there would be a case that one cannot calculate reliable waveforms. 
We find that one could modify the radiation reaction time 
by using the remaining gauge freedom in this class.
We propose the so-called radiation reaction gauge
where the self-force has only the radiation reaction component, 
which is derived by the radiation reaction formula. 
In this gauge condition, we find that 
we would have reliable waveforms of several years,
which is sufficient for the present LISA project.

Although we consider the radiation reaction formula
with an adiabatic linear metric perturbation 
may be sufficient for LISA project,
it remains a great theoretical challenge to calculate an orbit
longer than the radiation reaction time scale.

%%%%%%%%%%%%%%%%%%%%%%%%%%%%%%%%%%%%%%%%%%%%%%%%%%%%%%%%%%%
\section*{Acknowledgement}
%%%%%%%%%%%%%%%%%%%%%%%%%%%%%%%%%%%%%%%%%%%%%%%%%%%%%%%%%%%

YM thanks to Prof. Kip Thorne and Prof. Sterl Phinney
for encouragement.
YM was supported by NASA grant NAG5-12834
and NASA-ATP grant NNG04GK98G at CalTech.

%%%%%%%%%%%%%%%%%%%%%%%%%%%%%%%%%%%%%%%%%%%%%%%%%%%%%%%%%%%
%%%%%%%%%%%%%%%%%%%%%%%%%%%%%%%%%%%%%%%%%%%%%%%%%%%%%%%%%%%
%\appendix

%%%%%%%%%%%%%%%%%%%%%%%%%%%%%%%%%%%%%%%%%%%%%%%%%%%%%%%%%%
%\section{}
%%%%%%%%%%%%%%%%%%%%%%%%%%%%%%%%%%%%%%%%%%%%%%%%%%%%%%%%%%%

%%%%%%%%%%%%%%%%%%%%%%%%%%%%%%%%%%%%%%%%%%%%%%%%%%%%%%%%%%%

%%%%%%%%%%%%%%%%%%%%%%%%%%%%%%%%%%%%%%%%%%%%%%%%%%%%%%%%%%%
\end{document}